 \documentclass[twocolumn,british,pra, reprint]{revtex4}
 \usepackage[letterpaper]{geometry}
\usepackage{amssymb}

\usepackage{graphicx}
\usepackage[space]{grffile}
\usepackage{latexsym}
\usepackage{textcomp}
\usepackage{longtable}
\usepackage{multirow,booktabs}
\usepackage{amsfonts,amsmath,amssymb}
\usepackage{url}
\usepackage{hyperref}
\hypersetup{colorlinks=true,pdfborder={0 0 0},urlcolor=blue,citecolor=green,linkcolor=blue}
\usepackage[utf8]{inputenc}
\usepackage[english]{babel}



\begin{document}

\title{Augmented Reality with Hololens: Experiential Architectures Embedded in the Real World}

\author{Paul Hockett}
\affiliation{National Research Council of Canada, 100 Sussex Drive, Ottawa, K1A 0R6, Canada}

\author{Tim Ingleby}
\affiliation{Department of Architecture and Built Environment, Northumbria University, Ellison Place, Newcastle upon Tyne, NE1 8ST, UK}

\begin{abstract}
Early hands-on experiences with the Microsoft Hololens augmented/mixed reality device are reported and discussed, with a general aim of exploring basic 3D visualization. A range of usage cases are tested, including data visualization and immersive data spaces, in-situ visualization of 3D models and full scale architectural form visualization. Ultimately, the Hololens is found to provide a remarkable tool for moving from traditional visualization of 3D objects on a 2D screen, to fully experiential 3D visualizations embedded in the real world.

\begin{itemize}
\item Online version: \href{https://www.authorea.com/users/71114/articles/129932/_show_article}{Authorea}
\item Repository for videos and files, \href{http://dx.doi.org/10.6084/m9.figshare.c.3470907}{Figshare, DOI: 10.6084/m9.figshare.c.3470907}
\item This version submitted to \href{http://arxiv.org}{arxiv.org}, 13th Oct. 2016
\end{itemize}
\end{abstract}

\maketitle

\section{Introduction}

Augmented, or mixed, reality (AR or MR) has recently become easily accessible to a range of professionals for the first time, with the release of the developer edition of Microsoft's Hololens. Since April 2016, Microsoft has been supplying devices, albeit limited to developers in the US and Canada, whose application for the hardware was successful. At the time of writing (July/August 2016), a range of developers - from programmers to artists to scientists - are exploring this new hardware platform, developing applications and beginning to work seriously with AR. While AR platforms, and the underlying concepts, have been around for many years (as have related virtual reality (VR) technologies), Hololens represents a significant technological step, and the first opportunity for many professionals to experience head-mounted AR, and incorporate it into their everyday work.

As with many new technologies, the core applications and uses underlying AR are clear, and naturally formed some of the initial reasons for the hardware development (for recent discussions, within the context of architecture and design, see for instance \href{http://www.codessi.net/architecture-age-augmented-reality}{\textit{Architecture in an Age of Augmented Reality: Mobile AR's Opportunities and Obstacles in Design, Construction, and Post-Completion}}, ref. \hyperref[csl:1]{(Abboud, 2014)}, and \href{http://dx.doi.org/10.1016/j.compenvurbsys.2015.05.001}{\textit{To go where no man has gone before: Virtual reality in architecture, landscape architecture and environmental planning}}, ref. \hyperref[csl:2]{(Portman, Natapov, \& Fisher-Gewirtzman, 2015)}). For example, AR/MR presents an ideal environment for 3D visualization, in which existing real-world objects can be combined with computer-generated objects (c.f. VR, in which the environment is fully virtual with no real-world components). Initial promo and demo materials from Microsoft detail, and illustrate, some of these concepts, applied to CAD and architectural scenarios, e.g. refs. \hyperref[csl:3]{(Trimble, 2015)} \hyperref[csl:4]{(Microsoft, 2015)} \hyperref[csl:5]{(Lynn \& Erikson, 2016)}. However, also as with many new technologies, it is only when devices are released to a large development base (and, in the future, general users) that the full potential, depth and breadth of these basic concepts are realised. In the spirit of development, and cross-pollination, we detail herein some of our early work and experiences with the Hololens.

In this work, three basic usages are explored:
\begin{enumerate}
\item Basic multi-dimensional data visualization
\item Visualization of architectural forms: immersive and in-situ
\item Immersive data visualization: from data to computational architecture
\end{enumerate}
In these cases, user interaction is only at a basic level with a static 3D model; more advanced interaction will, naturally, be available in the future. Herein, our focus is on the workflow involved, the AR experience, including considerations of immersivity and scale of the experience, and our thoughts about the new platform. 

The work is primarily communicated in print via photographs, and online with additional video, data and 3D models available. These materials are referenced below where appropriate to the discussion herein; the full set of materials can be found online \href{https://dx.doi.org/10.6084/m9.figshare.c.3470907}{on Figshare} (ref. \hyperref[csl:6]{(Hockett \& Ingleby, 2016)}). In the electronic version of this article, videos hosted on Vimeo and Figshare are hyperlinked directly in the text; additionally, for print, videos are referenced conventionally, and listed by DOI at the end of the manuscript.

\section{Background}

\subsection{Interests}
One of us (PH) works in the physical sciences, with a general interest in data visualization, and a background which includes programming, CAD and (limited) 3D graphics development. One of us (TI), is an architect, with a general interest in visualization of architectural forms, and a background which includes 3D modelling and CAD. One interesting outcome of the information era, and the prevalence of computers as the major tool in many fields, is a significant overlap between professionals in very different spheres. In our case, there is significant overlap in many concepts of interest to us, although our expertise and approaches are very different. In particular, the use of data and programmatic methods in the physical sciences overlaps significantly with developments in computational/algorithmic architecture and generative arts: here AR presents a key bridging tool for realization of abstract forms and spaces as an experiential architecture embedded in the real world. Such experiential realisations would previously have been impossible in most cases during the design phase of a project. (For a related, and detailed, discussion on the emergence of interdisciplinary visualization with VR, see ref. \hyperref[csl:2]{(Portman, Natapov, \& Fisher-Gewirtzman, 2015)} and references therein.)

\subsection{Techniques}\label{sec:techniques}
In this work a range of computational methods were used. At this stage, the workflows are not yet streamlined or automated, nor are the 3D models visualized interactive, so these methods present only the most basic and readily accessible usages.
\begin{itemize}
\item For basic and immersive data visualization, pre-processing work was performed in \href{http://www.mathworks.com}{Matlab} (R2015b) and 3D data spaces were exported in a 3D XML format (.x3d) using the open-source tool \textit{Matlab 3D figure to 3D (X)HTML} \hyperref[csl:7]{(Kroon, 2011)}, available from \href{https://www.mathworks.com/matlabcentral/fileexchange/32207-matlab-3d-figure-to-3d--x-html}{the Matlab File Exchange, "fig2xhtml.m"}. The resulting files were piped through \href{https://www.blender.org}{Blender} to convert them to \href{http://www.autodesk.com/products/fbx/overview}{Autodesk Motionbuilder format (.fbx)}, which is currently the preferred 3D format supported by the Hololens for basic 3D viewing. 
\item For visualization of architectural forms, original work from \href{http://www.sketchup.com}{Google Sketch-Up (.skp format)} was exported to \href{https://www.khronos.org/collada/}{Collada format (.dae)}, and again piped through Blender for .fbx conversion.
\item For computational architectural forms, original work was created in \href{https://www.rhino3d.com}{Rhinocerous 5} with the \href{https://www.grasshopper3d.com}{Grasshopper plug-in}, and imported into a \href{https://unity3d.com/partners/windows/hololens}{Unity Hololens project}. The project was then compiled and deployed onto the Hololens via \href{https://www.visualstudio.com}{Microsoft Visual Studio 2016}. This workflow is more involved, but is the most general methodology, and allows for interactive and programmatic content to be developed within Unity and Visual Studio.
\end{itemize}

\begin{figure}[h!]
\begin{center}
\includegraphics[width=0.7\columnwidth]{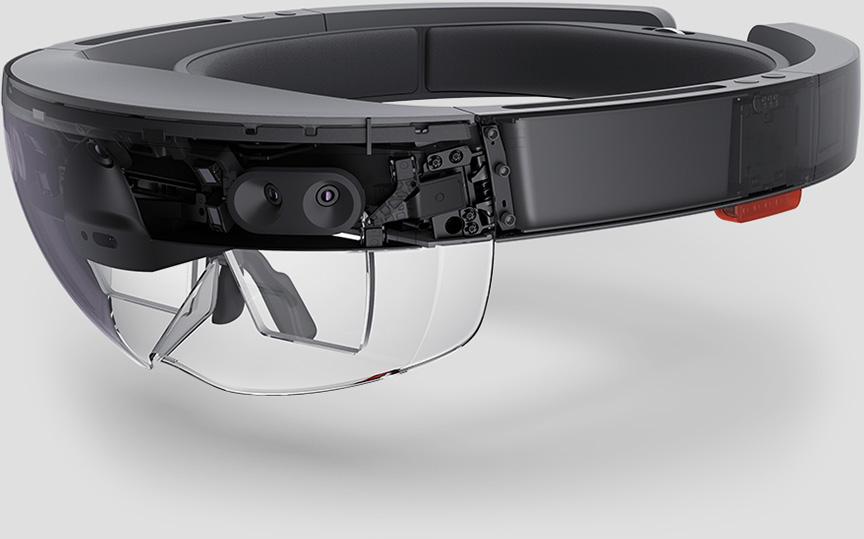}
\caption{{\label{fig:hololensIntro} The Microsoft Hololens for AR/MR (figure reproduced from \href{https://developer.microsoft.com/en-us/windows/holographic/hardware_details}{Microsoft Hololens hardware webpage}).%
}}
\end{center}
\end{figure}

\begin{figure}[h!]
\begin{center}
\includegraphics[width=1\columnwidth]{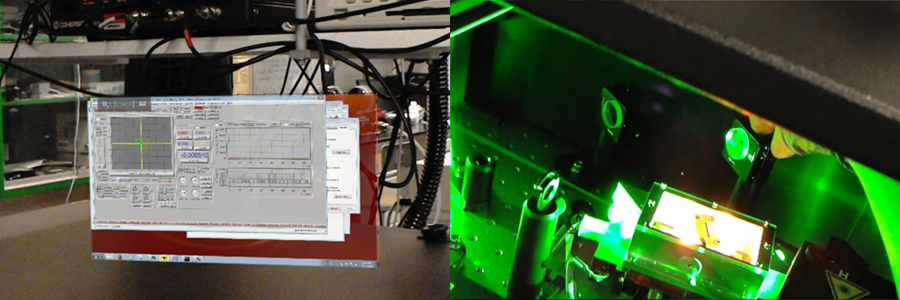}
\caption{{\label{fig:basicPOV} A basic MR environment from the user PoV showing, in this case, a remote-desktop feed projected into real space (left panel). Here the usage was in a laboratory environment, with the instrument feed utilized while work was performed on a laser system (right panel). Further details on this usage case can be found in the source video \href{http://femtolab.ca/?p=712}{\textit{Hololens @femtolab.ca: week 4, basic laser lab use}} (ref. \hyperref[csl:9]{(Hockett, 2016)}).%
}}
\end{center}
\end{figure}

The Hololens is, essentially, a head-mounted Windows 10 device (see figure \ref{fig:hololensIntro}). The display is a translucent visor-based system, allowing the user to view both real and virtual worlds (see figure \ref{fig:basicPOV}). Interaction is via gaze, gesture and voice inputs or, optionally, wireless peripherals. The Hololens is currently available to US and Canadian developers (as of late April 2016); a commercial release date has not yet been announced. Further details of the hardware platform, and other information, can be found on the \href{http://hololens.com}{Hololens website}, ref. \hyperref[csl:8]{(Microsoft, 2016)}.

With respect to AR visualization (and in comparison with existing VR hardware platforms), some of the key features of the Hololens are:
\begin{itemize}
\item Fast and accurate spatial mapping, via an IR ``depth" camera and visible light cameras.
\item Dual HD translucent display.
\item Integrated unit for untethered operation: no cables or support hardware required.
\item Gesture and voice command recognition.
\end{itemize}

These features, in particular, separate the Hololens from existing AR and VR hardware, and provide an excellent platform for AR visualization in any environment, with the user completely free to move.

\section{Case Studies}

\subsection{Multi-dimensional data visualization}\label{sec:basic-data-vis}
\begin{figure}[h!]
\begin{center}
\includegraphics[width=1\columnwidth]{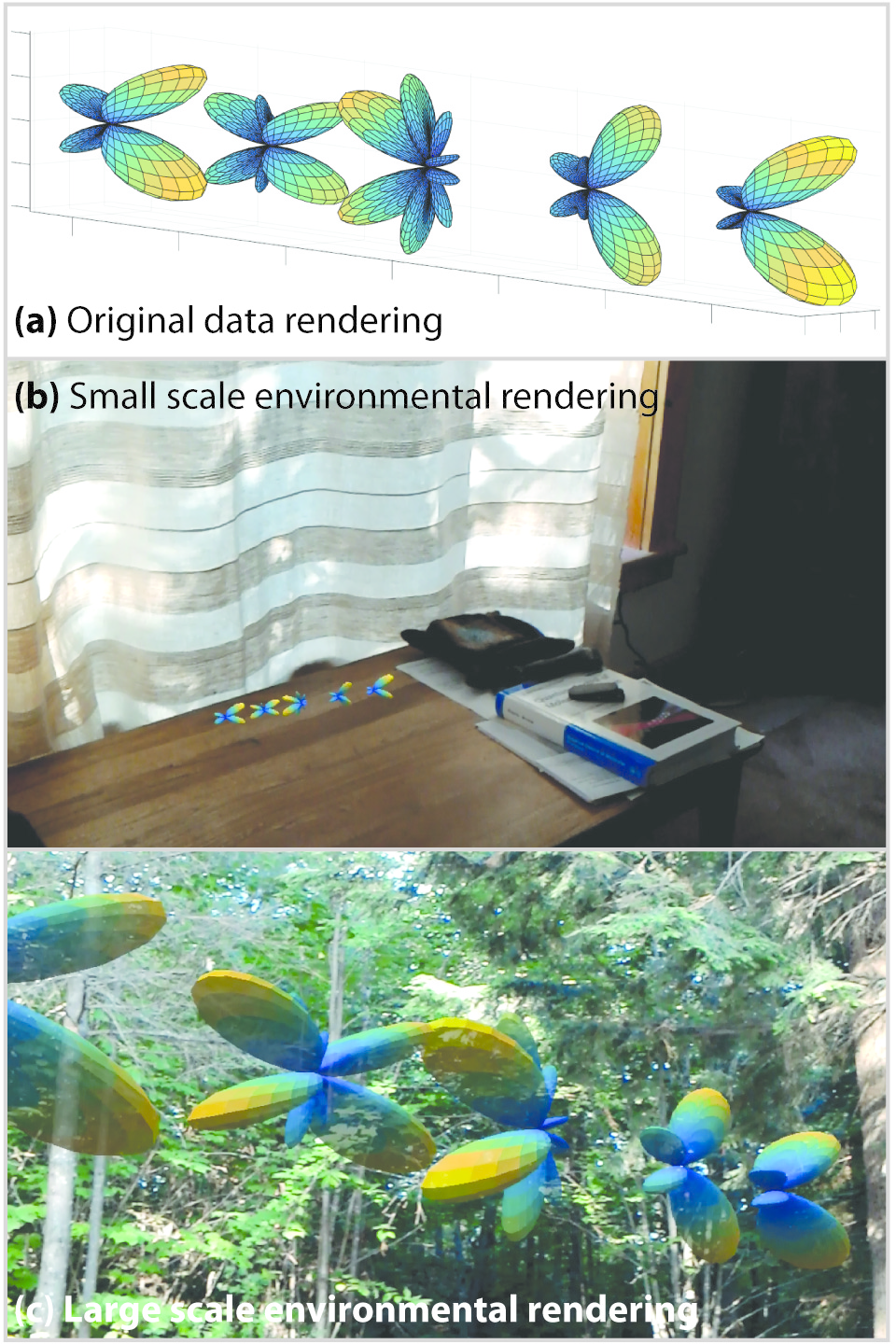}
\caption{{\label{fig:cs2_data} Data visualization of quantum mechanical scattering calculations. (a) Original data visualization. (b) \& (c) Stills from video footage recorded on the Hololens, showing the data visualization in a mixed reality environment at small (tabletop) and large (sculptural) scales. Video material, \textit{Hololens @femtolab.ca: week 2, Basic data visualization and immersion}, can be found online on \href{https://vimeo.com/175310662}{Vimeo} or \href{http://dx.doi.org/10.6084/m9.figshare.3842907}{Figshare} (ref. \hyperref[csl:10]{(Hockett, 2016)}).%
}}
\end{center}
\end{figure}

The first usage case is ``basic" AR data visualization. In this capacity, data is visualized in the AR/MR environment, and the user is free to explore and interact with the 3D forms. Figure \ref{fig:cs2_data} shows examples of this basic usage; a better impression of the user experience and capabilities of the Hololens can be gained via video footage, \textit{Hololens @femtolab.ca: week 2, Basic data visualization and immersion}, which can be found online via \href{https://vimeo.com/175310662}{Vimeo} or \href{http://dx.doi.org/10.6084/m9.figshare.3842907}{Figshare} (ref. \hyperref[csl:10]{(Hockett, 2016)}). The data shown here comprises computational results showing quantum-mechanical electron scattering in the molecular frame, for a CS$_{2}$ molecule undergoing vibronic dynamics on the femtosecond timescale - see ref. \hyperref[csl:11]{(Hockett, Bisgaard, Clarkin, \& Stolow, 2011)} for further details. Each surface visualized in the demonstration is the result of a calculation for a different ionization time $t$, ranging from 100 to 900 femtoseconds, with the form of the distributions changing as the molecule bends and stretches.

In terms of visualization, these distributions manifest as sculptural forms and present a pleasing aesthetic. Each object is a surface $I(\theta,\phi; E,t)$ that can be described by $I(\theta,\phi; E,t)=\sum_{L,M}d_{L,M}(E,t)Y_{L,M}(\theta,\phi)$ - essentially an expansion in standard spherical harmonic functions $Y_{L,M}(\theta,\phi)$ - and plotted in spherical polar coordinates $(|I|,\theta,\phi)$. In the visualizations shown here, the colour-mapping also shows the magnitude $|I|$. The symmetry and form of the distributions is correlated with the molecular geometry, and the underlying quantum mechanical scattering process which determines the values of the expansion parameters $d_{L,M}(E,t)$. Here $E$, the scattering energy, is fixed over all the distributions. For the Hololens user, these forms present a visceral way to explore and understand such results. While only a few results are shown here, the Hololens could be used to visualize a much larger data set, thus allowing for rapid appreciation and comparison of results in a natural 3D space, in a way not easily achieved on a computer screen.

In other ways, this type of visualization does not explore the full potentials of the Hololens, since these surfaces are shown as only a function of two variables, and are hollow. The computer-generated aspects of the visualization are situated in the real-world, but do not interact with it or ``require" it. Although the user experience is visceral, and more flexible than on a computer monitor in terms of the size of the data space visualized, insight into the results themselves is (arguably) not significantly enhanced in this particular case. However, from both a technical and a purely artistic perspective, AR visualization of even such relatively simple data forms is quite beautiful and remarkable. One can readily imagine the possibilities here for encouraging and enhancing overlap between different fields, the presentation of scientific results, communication of complex forms with other researchers or non-specialists at a visceral level, interactive presentation with remote users, and so on and so forth. Additionally, it is interesting to experience navigation of the data space in a ``natural", physical way - via gaze, gesture and motion - rather than through the use of intermediary tools (mouse, keyboard etc.). Surprisingly, this interaction feels natural and immersive, and not at all inconvenient, tiring or difficult, although other users may feel differently.

\subsection{Visualization of architectural forms}

The second usage case is one where the potential application of AR is more directly immediate.  The ability to conceptualise, visualise and communicate 3-dimensional ideas to both oneself and non- architecturally-educated audience has long been a challenge for architects.  Historically the development of drawing techniques, most notably isometric and perspective projections together with physical scale-models, have helped address this. Such modes of representation are however necessarily static and rely upon scale-clues to help viewers situate themselves and/or contextualise proposals.  Since the latter part of the 20th Century evolving CAD technologies have `solved' the issue of stasis, firstly through `walkthroughs' or `flythroughs' of 3-dimensional space (animated sequences following a pre-determined path), and latterly through VR objects, panoramas (360 degree circular or spherical navigable views) and most recently real-time rendering - for a recent discussion in the AR sphere, see \hyperref[csl:12]{(Schubert, Schattel, Tönnis, Klinker, \& Petzold, 2015)} and refs. therein.  Nevertheless, the issue of scale persists.  It is here, in the visualisation of a 3D model in an immersive way at true-life scale, that AR suggests itself as a significant addition to the tools at an architect's disposal, both in the design and communication of architectural ideas, allowing for the architect to explore their ideas (and others to engage with them) in different ways.  

The application considered here is the design for a Trekking Cabin to be located in Iceland.  The design process used may be considered `traditional' in the sense that the design went through a series of design iterations that enabled the design to be increasingly refined and resolved in relation to the set architectural brief.  The design methods may also be considered to be (largely) traditional in the sense that they employ widely-used analogue and digital drawing and modelling techniques as part of the design process, culminating in a series of presentation drawings illustrating the finalised proposal as 2D- and 3D-representations.  

\begin{figure}[h!]
\begin{center}
\includegraphics[width=1\columnwidth]{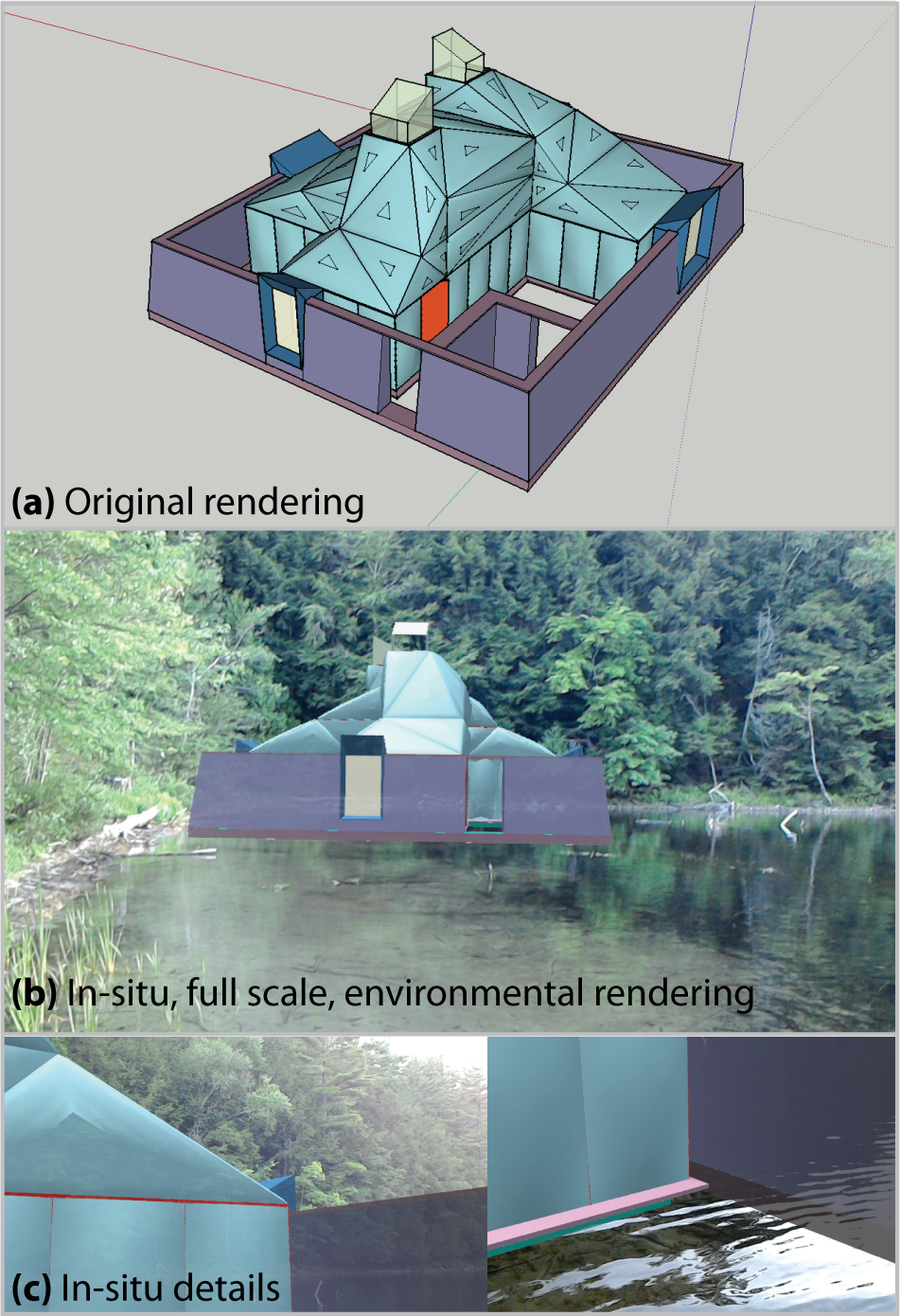}
\caption{{\label{fig:cabin-vis} Architectural visualisation at full scale \& in-situ. (a) Original rendering of the trekking cabin design (Sketch-Up). (b) Full scale, in-situ rendering. (c) In-situ details, showing mixed-reality aspects of the visualisation in the courtyard of the cabin. Video material, \textit{Hololens @femtolab.ca: week 3, Large Scale and In-Situ Visualizations}, can be found online via \href{https://vimeo.com/184204010}{Vimeo} or \href{http://dx.doi.org/10.6084/m9.figshare.3846672}{Figshare} (ref. \hyperref[csl:13]{(Hockett \& Ingleby, 2016)}).%
}}
\end{center}
\end{figure}

Less conventionally, nascent Hololens AR (as described in sect. \ref{sec:techniques}) was used as part of the design development process.  This case study is illustrated in figure \ref{fig:cabin-vis}, and video material, \textit{Hololens @femtolab.ca: week 3, Large Scale and In-Situ Visualizations}, can be found online via \href{https://vimeo.com/184204010}{Vimeo} or \href{http://dx.doi.org/10.6084/m9.figshare.3846672}{Figshare} (ref. \hyperref[csl:13]{(Hockett \& Ingleby, 2016)}).  AR undoubtedly has potential for both the development and presentation of architectural form, and both aspects shall be duly considered; however the use in this instance was governed by a defined submission requirement for the design and, of course, the currently limited availability of the AR technology.  Aspects of such Hololens usage have been considered already, although it is noted these tend generally to be in the form of promotional videos, see for instance refs. \hyperref[csl:5]{(Lynn \& Erikson, 2016)} and \hyperref[csl:4]{(Microsoft, 2015)}.

The first experience of architectural form by means of AR at true-life scale (or as close as the available space allows) is certainly striking. The ability for the user to navigate freely around and through a design creates something akin to a `sandbox' environment and while such freedom of exploration may be considered equally achievable by means of real-time renderings it is significantly distinguished by both how space is perceived when immersed within the AR environment and by the haptic experience of such.  What is also striking, and potentially of great potential for the architect as designer, is how the design itself can be immersed or synthesised within a real-life environment.  It is easy to envisage that testing a design against a specific site would enable designs to be accurately adjusted in response to their immediate context as part of an iterative process, or indeed to imagine new workflows and design processes emerging as a result of such `live' testing. It should be noted that the Trekking Cabin is intended as a prototypical design that can be adapted to respond to a range of sites, in this instance therefore the testing of it within an `alien' environment (i.e. Canada as opposed to Iceland) was no less beneficial to the design development process.

The same things that make AR so compelling for the visualisation of architectural forms are also, perhaps, the notes for greatest caution at this point in time.  The Trekking Cabin is a small-scale building on a single level and intended for an open site and this makes finding a space to experience it in relatively simple.  Any design of greater scale or sectional complexity would of course require greater space or other means (treadmills or peripheral controls) to navigate around and between levels, this may somewhat diminish the haptic experience.  Moreover, the complex urban environments that the vast majority of development takes place in today rarely offer such virgin site conditions as the Trekking Cabin is intended to occupy, and this may prove restrictive to the potential for feedback through in-situ testing.  

It is also important in this consideration to make a distinction between architectural design and architectural visualisation, the latter of which has moved unerringly from representational to photo-realistic modes as computing power has increased.  This is significant as it establishes a certain level of expectation about how we consume architectural visualisations.  While designing, it is common for an architect to consciously or sub-consciously hold a large amount of design information within their head.  It is also not uncommon that aspects of the design will remain ambiguous or unresolved (at least until much later in the process), usually quite deliberately: representational visualisation is well-suited to this. Some such `gaps' are apparent within the Trekking Cabin development study model seen in the video: certain geometries are unresolved while the model's lurid diagrammatc colour scheme serve to inadequately represent the true material or light qualities of the space - the interaction between which is an essential architectural quality.  In terms of the design process, workflows and technology, this potential gap may be initially difficult (and to the architect, even undesirable) to bridge, not least because the processing power required to achieve a satisfactory level of heightened realism, not so much for material mapping but certainly with regard to complex lighting physics of a space (even baked in), is almost unquantifiable at this point.

Aspects of this issue have been discussed in depth, in the context of AR for architecture, by Schubert et. al. \hyperref[csl:12]{(Schubert, Schattel, Tönnis, Klinker, \& Petzold, 2015)}, as part of their recent work developing interactive AR tools. The more general philosophical considerations of \textit{representational realities}, in the context of early VR, were also sketched succinctly by Gregory:

\begin{quote}
A drawing may represent truth or falsity, generally a complex mixture. Writing may simply miss out details which inadvertently are wrong in drawings, for gaps are better tolerated in prose than pictures, so prose can be less misleading. We are used to this, and much of scholarship is challenging truth of other scholars and how they present their facts and ideas. This involves making explicit what is pictured or conveyed, producing lively and often fruitful debate. Here there are rules and politenesses that preserve the vitality and integrity of scholarship. But it takes time---many years---to get into this. It is not available for children, or the uninitiated learning science. It is here that VR partial-yet-seeming-true pictures seem most dangerous. Yet it is just here that VR is most likely to be used and trusted. \hyperref[csl:14]{(Gregory, 1995)}
\end{quote}

\subsection{Immersive data visualization: from data to computational architecture}

Naturally following from the above ``basic" usage cases, AR visualization can be conceived as a bridge between data-driven and anthrophic forms or design. In the former case, the use of AR provides a way to scale-up data spaces to an experiential size, and immerse the user in a data space from any source; in the latter case, the use of data-based design - making use of either pure data, or a generative design processes, as in computational architecture - combined with large scale and/or site-specific experiential visualization, clearly provides a route from more traditional design practice.

In this work, both approaches have been trialed.  In the former case, a volumetric data space was visualized as first a sculptural form, then as an immersive data architecture; in the latter case, basic generative computational design was performed, then visualized at full scale in an MR environment.

\subsubsection{Immersive data to architectural form}
\begin{figure}[h!]
\begin{center}
\includegraphics[width=1\columnwidth]{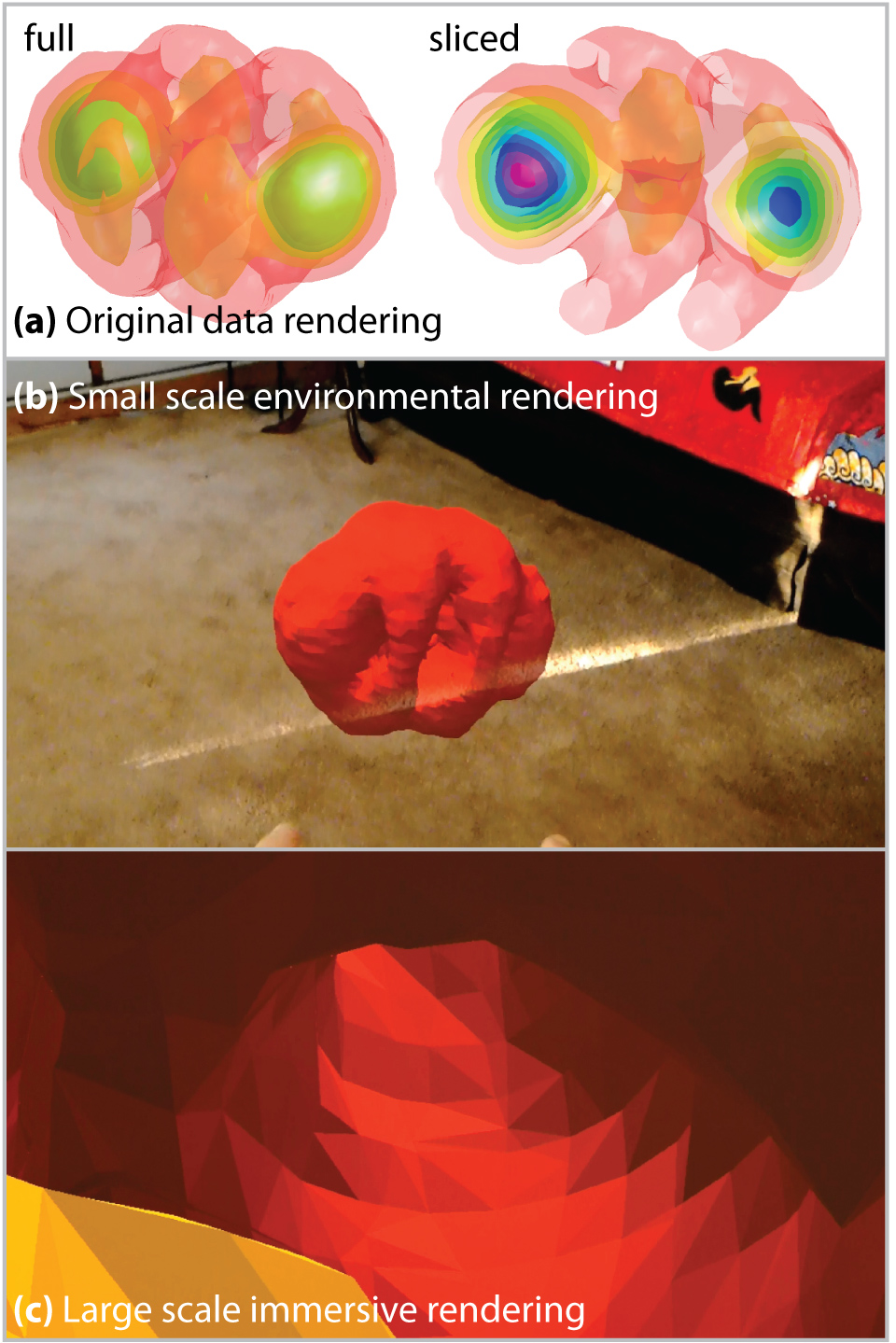}
\caption{{\label{fig:tomo_data} Data visualization of volumetric electron scattering data. (a) Original data visualization, for the full data space with translucent isosurfaces. (b) \& (c) Stills from video footage recorded on the Hololens, showing the data visualization in a mixed reality environment at small (tabletop) and large (sculptural) scales. Video material can be found online on \href{https://vimeo.com/175310662}{Vimeo} or \href{http://dx.doi.org/10.6084/m9.figshare.3842907}{Figshare} (ref. \hyperref[csl:10]{(Hockett, 2016)}).%
}}
\end{center}
\end{figure}

For an immersive data space example, experimental 3D electron scattering data $I(E,\theta,\phi)$ was used. This data is similar to the computational results used in sect. \ref{sec:basic-data-vis} but is, in this case, true volumetric data with three variables, defined in a 251x251x251 voxel space. Renderings and photographs from the Hololens are shown in figure \ref{fig:tomo_data}, and video material is again available online on \href{https://vimeo.com/175310662}{Vimeo} or \href{http://dx.doi.org/10.6084/m9.figshare.3842907}{Figshare} (ref. \hyperref[csl:10]{(Hockett, 2016)}). Scientific and technical details are available in ref. \hyperref[csl:15]{(Hockett, Lux, Wollenhaupt, \& Baumert, 2015)}, while further aspects of the scientific visualization and exploration of such data (and difficulties thereof, especially in higher-dimensional data spaces) can be found in ref. \hyperref[csl:16]{(Hockett, Ripani, Rytwinski, \& Stolow, 2013)}.

The user experience was, in this case, surprisingly immersive despite the low resolution data environment.  The data space representation used a standard colourmap (HSV), but the use of only 10 bounding isosurfaces (originally selected for clarity in static form, i.e. for journal articles in print or on-screen) led to a definite "8-bit" feel to the space, reminiscent of early computer graphics, and early artistic visualizations of immersive environments such as those shown in the films Tron (1982) and The Lawnmower Man (1992). At an immersive scale this reductive colour-mapping became quite psychedelic, adding to the experience.

As a tool for pure multi-dimensional data visualization and exploration, VR is likely a better choice (for a more immersive user experience, and the ability to add more processing power at the back end); however, AR is likely to prove more convenient in many usage scenarios, such as in the lab or in the field.  While these scenarios do not make use of any aspects of the real-world environment, they do benefit from the portability of the Hololens.

As a tool for transformational visualization, however, the Hololens is quite unique, with the ability to experience a data space not just as an immersive form, but as an immersive form embedded in the real-world, thus transforming the abstract into an experiential architecture in the real world. This tool clearly opens new avenues for any practitioner/researcher interested in the use of abstract computational spaces in architecture, or other experiential installations (e.g. large-scale art installations).

\subsubsection{Computational architecture to the real-world}
Our final useage case provides a brief exploration of visualization of forms established by standard computational architecture tools, in this case Rhino and Grasshopper. While conceptually quite different to the preceding discussion, technically this exploration is very similar, with only the initial source of the visualized spaces changed. However, the direct response of the 3D architectural forms to user input in this type of design process is one which is very suggestive of the future power of AR. In such generative cases, modifying parameters to the driving equations, changing the sketches which form the skeleton upon which the equations operate, or changing the equations themselves, can all cause significant and radical changes to the final design. Here, the hand of the architect is in the creation of this complex generative environment, and in the final selection of an aesthetically pleasing generative result, but the architect is not a direct agent in the creation process itself.

One direct consequence of this type of design process: a multitude of possible forms may be generated very rapidly, and interactively, from even the simplest underlying equations. Any tool which allows for an experiential visualization of such structures, and places them in the environment within which they may, ultimately, be physically constructed, is therefore of great utility for this design philosophy. Other possible enhancements to a traditional generative design methodology and practice include: the practical aspects of easily visualizing a large selection of the forms created; the artistic aspects of appreciating the generated forms sculpturally, in a fully 3D environment; the architectural experience and ability to create the forms as immersive environments, at scale, or in specific real-world locations.

As discussed above, in these initial explorations the Hololens was used only for visualization and basic manipulation, with no direct feedback from the user to allow for changes in the visualized form. Here we illustrate just a static example of a generative architecture. However, in principle, there should be no issue with providing such feedback, given time to develop suitable applications: early concept examples can be found in the Hololens demo materials \hyperref[csl:4]{(Microsoft, 2015)} \hyperref[csl:3]{(Trimble, 2015)} and, since release, demos from independent developers are beginning to appear \hyperref[csl:17]{(Brekelmans \& Brekel.com, 2016)}. In this case, feedback would clearly be of great benefit, and allow for rapid changes in the computational form as mentioned above. The workflow in this case made use of a fully expandable and programmatic methodology, using Unity and Visual Studio (as detailed in sec. \ref{sec:techniques}), which is the generic workflow for creating ``holographic" applications for the Hololens. Hence this usage case illustrates the seed of more complex and interactive 3D experiences.

\begin{figure}[h!]
\begin{center}
\includegraphics[width=0.9\columnwidth]{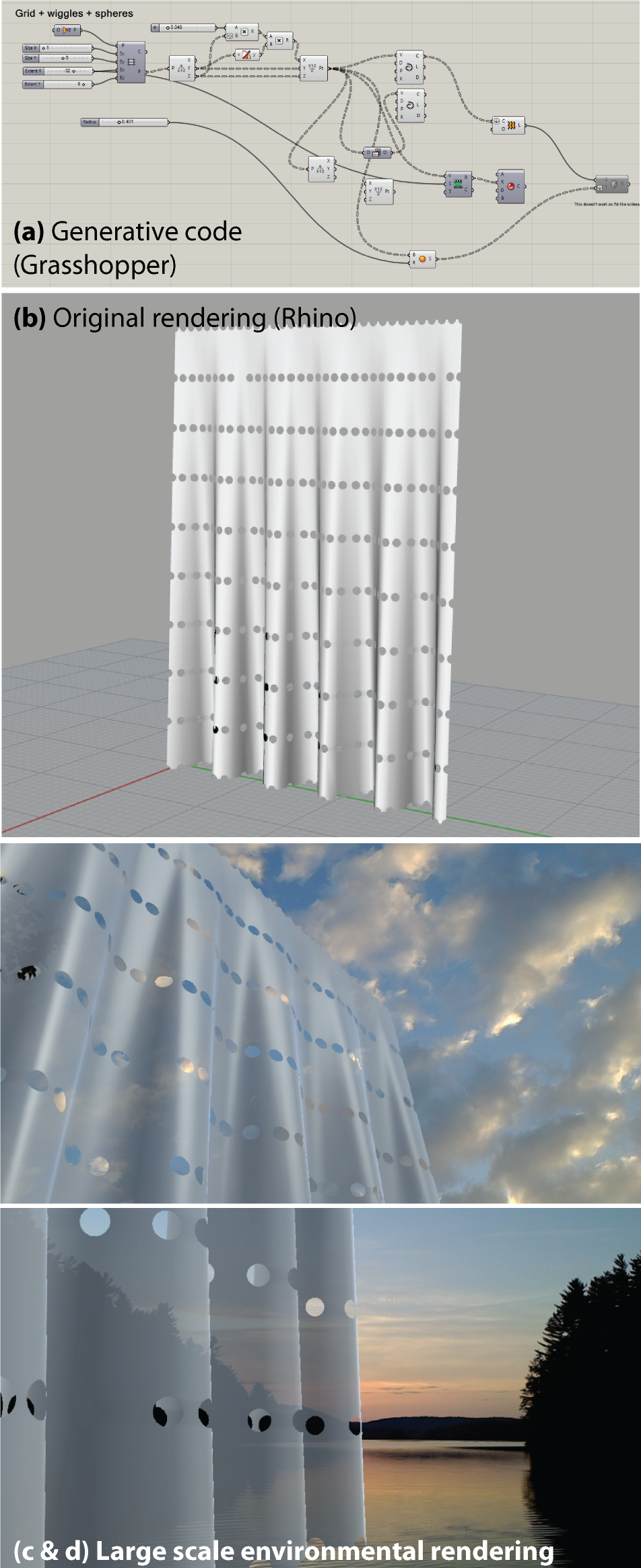}
\caption{{\label{fig:gen-arch}Illustration of computational architecture using Rhino and Grasshopper (top images), and visualized on the Hololens (lower images). The Rhino rendering is at metre scale, 1 unit square equals 1 m. Video of this case study, \textit{Hololens @femtolab.ca: Computational Architecture at Large Scales}, can be found online on \href{https://vimeo.com/182387356}{Vimeo} or \href{http://dx.doi.org/10.6084/m9.figshare.3846675}{Figshare} (ref. \hyperref[csl:18]{(Hockett \& Ingleby, 2016)}).%
}}
\end{center}
\end{figure}

The case study is illustrated in figure \ref{fig:gen-arch}, and video material, \textit{Hololens @femtolab.ca: Computational Architecture at Large Scales}, can be found online on \href{https://vimeo.com/182387356}{Vimeo} or \href{http://dx.doi.org/10.6084/m9.figshare.3846675}{Figshare} (ref. \hyperref[csl:18]{(Hockett \& Ingleby, 2016)}). In this example a large, grid-based structure was created, with sinusoidal surface deformations applied as a function of height. Circular cuts were also made, following the same sinusoidal patterning. The surface created was deliberately large, with dimensions of approximately 10x2x30 m (length x depth x height), providing an interesting test of the Hololens for the rendering of large objects.

The full-scale visualization necessarily requires a large space in order to experience, and was placed at the edge of a lake in the study illustrated here. In this case, the scale of the visualization was quite convincing, and the user found the experience immersive. The beauty of the real-world location, and late-evening light, added significantly to the mixed-reality scene, and brought a serendipitous temporal and site-specific aspect to the visualization. Watching the clouds roll past this virtual form was both impressive and remarkably realistic, even with the partial transparency and textural flatness of the architecture in this particular representational reality.

\section{Discussion, Conclusions and Future Directions}

As discussed by Wolfgang and Hentschel in \textit{Quo Vadis CAVE: Does Immersive Visualization Still Matter?} \hyperref[csl:19]{(Kuhlen \& Hentschel, 2014)}, ``scientific visualization had been promulgated as a killer app for VR technology almost from day one"; within certain contexts the same statement clearly applies to AR/MR. In the same vein, the authors also discuss the specific advantage of immersive data visualization, and ``natural" spatial interactions:

\begin{quote}
 ...we argue that the key advantage of immersive visualization stems from natural, spatial interaction. This includes a range of techniques --- for example, head tracking and view-dependent projection --- that let users intuitively shift their viewpoint. It also includes 3D interaction metaphors that let users control a variety of visualization parameters from within their data. To be more specific, we believe that the potential of immersive visualization is about neither the raw pixel count
nor other display-centric characteristics. It's about users being immersed in their data. \hyperref[csl:19]{(Kuhlen \& Hentschel, 2014)}
\end{quote}

As discussed above, the immersive and (surprisingly) natural feel of data visualization with the Hololens - even for simple volumetric data space renderings - fulfills these criteria. While our current basic case studies do not reach as far as interactivity, the power and promise of the tool is already clear here. From this point, we are working towards the development of dedicated applications for data exploration, with a range of possibilities envisaged. Desirable features for such an application including making full use of the portability of the Hololens for in-situ and on-the-fly use (for example, in the laboratory or on the building site), and interactivity in the type, style and layout of the visualization. Including a fourth dimension to data structures, which could be visualized either spatially via discrete 3D spaces/objects, or in a time-dependent (animated) manner, is another obvious possibility.  In general, these developments would allow for a user-focussed workflow and real data exploration far beyond that provided by static visualizations.

While immersive data visualization is certainly a promising avenue of use for the Hololens, the architectural case studies herein illustrate a more obvious usage case which can immediately be appreciated and intuitively understood; this type of usage is likely to appeal to a wider user-base of both professionals and non-professionals alike. In the case of architectural visualization, the natural feel of the environment is paramount, since the representational reality seeks to simulate actual reality - as opposed to an abstract multi-dimensional data space. Although our case studies did not make use of particularly detailed or sophisticated renderings, thus were still somewhat abstract in form, the architectures trialed herein were already convincing, immersive and visceral for the user, and readily fulfilled the promise of AR/MR for experiential architectures embedded in the real world despite the ``representional" nature of our forms. In future work, we anticipate the use of more detailed 3D environments and closer integration with the real world environment, as well as interactivity in the environment, will all play an important role. All these aspects would further the basic usages of the Hololens illustrated herein, which only just begin to scratch the surface of the possible, and would certainly provide powerful tools for architectural design, visualization and communication, and the possibility for furthering interdisciplinary uses. As noted by Schubert et. al., in the context of levels of representional realities and AR tools for design and presentation, ``... we need to progress beyond the current rigid methods to facilitate a direct and intuitive way of working with the design idea. Only then will it be possible to directly develop and interactively present design ideas for all participants whether experts or laypeople." \hyperref[csl:12]{(Schubert, Schattel, Tönnis, Klinker, \& Petzold, 2015)}.

In terms of the user experience, large scale visualization pushes the capabilities of the Hololens. In the cases here, two aspects of the hardware are quite apparent to the user during the experience, although may not be clear in the recorded experience. 
\begin{enumerate}
\item The spatial position of the 3D object fluctuates somewhat.
\item The FoV of the Hololens (which does not fill the user's perhipheral vision) is quite apparent for large objects.
\end{enumerate}
Neither issue is major and, given the impressive spatial mapping of the platform it seems churlish to raise this particular issue. However, in the worst case, these aspects may take the user out of the immersive experience. It is of note that these issues are extremely minor in general usage - for example when visualizing objects at smaller scales or using Hololens-based Windows applications (essentially 2D virtual screens placed in the environment). In our particular case study, the spatial mapping did lose track of the environment due to the lack of recognisable objects for the visible light cameras and/or the inherent range of the ``depth camera" ($\approx$\href{https://developer.microsoft.com/en-us/windows/holographic/spatial_mapping_design}{3 m}) - but operating in a wide-open area, with a fluid and moving floor, is clearly an extreme usage case for the Hololens. Nonetheless, aside from occasional glitches and shifts of the visualization, even in this case the spatial stability was sufficient to be convincing.

One reason to raise these issues is due to the \textit{expectation gap} of any new technology. This is oft-cited as one of the reasons for the collapse of VR in the mid-1990s, and refers to the potential for professional users (or, ultimately, consumers) to abandon a platform or, in extremis, a whole technology, if early promise is not met \hyperref[csl:20]{(Bright, 2015)}. This may be caused or exacerbated by a range of factors, including hyperbole in advertising, a poor user experience, and unreliable technology. Obviously, with immersive platforms, the expectation gap may be particularly severe, and the penalty for failure similarly so, particularly when devices have the ability to cause physiological problems in users. Some have cited this as the main reason for the very gradual roll-out of recent wearable and immersive platforms, e.g. Google Glass, Oculus Rift and the Hololens, with the number of devices limited, and with developers/researchers as the target market. These types of issues in technology development are covered succinctly by \href{http://www.gartner.com/technology/research/methodologies/hype-cycle.jsp}{Gartner's Hype Cycle}. Within this framework VR/AR has long resided in the ``trough of disillusionment", but one might consider the current generation of AR/VR to now be on the ``slope of enlightenment" towards a usable, useful and productive technology platform.

In this case, for our uses, and based on the state of technology in 2016, the Hololens lives up to our expectations in most areas, and exceeds in some. Initial experiences of AR environments using the Hololens were immediately compelling, although working with the technology over a more extended period revealed glitches as discussed above that, while relatively minor, can be frustrating or diminish the immersivity of the experience. However, others' expectations and usage cases may differ (see, for instance, \href{http://arstechnica.com/gadgets/2015/07/the-hololenss-limited-field-of-view-doesnt-matter-and-heres-why/}{\textit{The HoloLens' limited field of view doesn't matter, and here's why - It's all about who you sell it to}} \hyperref[csl:20]{(Bright, 2015)}) - in the current form the graphical capability of the Hololens will not match high-end VR for example, or dedicated gaming machines. For example, much of the press coverage of the Hololens to date has focussed on the FoV (see, for a typical example, \href{http://arstechnica.com/gadgets/2015/05/hololens-still-magical-but-with-the-ugly-taint-of-reality/}{\textit{HoloLens: Still magical, but with the ugly taint of reality}} \hyperref[csl:21]{(Bright, 2015)}). While it is, of course, natural to raise both the pros and cons of a new hardware platform, contextually these comparisons may be somewhat fallacious and, additionally, often willfully miss all that is incredible about the technology, not to mention ignore what is, one presumes, not technically and/or economically feasible at the current time \hyperref[csl:22]{(Geng, 2013)}... but the expectation gap is unlikely to be reasoned with in this fashion.

Despite these caveats, the cases discussed here demonstrate that, although still in its infancy, AR provides considerable new possibilities for visualizing and experiencing information (for recent discussions see, for example, refs. \hyperref[csl:23]{(Olshannikova, Ometov, Koucheryavy, \& Olsson, 2015)} \hyperref[csl:24]{(Kwon, Muelder, Lee, \& Ma, 2016)} and references therein). It is anticipated that AR platforms such as Hololens and others will become widely adopted and ultimately commonplace, not just because of the relative affordability of the technology, but also because of the natural feel of the experience: this will only increase over time as the hardware and software evolves, as interfaces are made more intuitive, and as the issue of how large-scale AR environments (in terms of either the physical or data-set size), can be made more readily navigable, are addressed. As such AR offers great potential both for those who generate data or information and for those who consume it.

In a world in which the consumption of information has become increasingly through visual means, perhaps the primary significance of development of AR -- be it for physics or architecture as discussed here, or many other fields that traditionally rely on having \textit{a priori} knowledge or skills of a subject to be easily engaged with, is the potential for information to become more widely accessible (and consumable) and for users to actively engage with, even in the absence of \textit{a priori} knowledge.  This ability to interact suggests AR could provide a means by which fields of knowledge or subjects that were previously conceived to be 'closed', may become increasingly democratized (and thus given greater currency?).

So, if AR does present just such an opportunity, how might 'closed' fields of study capitalize?  After all, it is easy to see how new technologies' efficacy can be easily compromised by novelty (not that the two need necessarily be mutually exclusive).  Perhaps as with any tool, one answer may lie in the intelligence with which it is used or applied, for which the cases presented here suggest two alternative strands.  

In the first strand the AR environment is presented in a form that is comprehensible but fixed, the end user's engagement with it is therefore (relatively) passive.  How engaging any information presented in this way is relies largely on how compelling it (or the presentation of it) is.  As alluded to earlier, the visualization of information has become an increasingly specialized field and with even traditionally visually-literate professions such as architecture now (willingly or unwillingly) ceding some responsibilities to such specialists it is eminently probable that this trend will continue.  However, in divesting authors of such responsibilities, the level of control exerted is diminished accordingly, perhaps all the more so in less visually-oriented disciplines.  If for no reason other than the limitations in processing power, the short- to mid-term future of AR appears likely to lie in more representational forms as anticipated here, the advent of AR marks a moment in time when authors, emancipated by an abundance of ever-more accessible software and the diminished demands of representational visualization, can reclaim control.   Of course this alone is unlikely to result in a compelling end-user experience, so perhaps one future lies in AR enabling authors of information to have more of an interlocutory role in its interpretation and subsequent visualization.

In the second strand the AR environment can be not only experienced but also (re-)shaped, in actively engaging the user's role is elevated from that of consumer to protagonist.  Moreover, to create such an experience it is necessary to redefine the role of architect of the AR environment from that of a 'master builder' that creates a single optimum solution to a facilitator who uses their broad understanding of a context to establish a set of rules or parameters within which there are a multitude of possible 'valid' outcomes. (It is of note that there is significant precedent here in generative arts in general, and the recent evolution of procedural or generative computer games in particular, e.g. \href{https://en.wikipedia.org/wiki/No_Man's_Sky}{No Man's Sky}.)  This redefinition of roles will undoubtedly be perceived by some as a threat, and certainly there is an argument that this could result in a ceding of yet further control, albeit to potentially different agents.  To others it may seem the inevitable next (positive) step in the democratization of information. Any speculation as to the qualitative outcomes of such a move is at this stage a result of the inevitable conjecture that emerges in the absence of anything more concrete to base judgement on; as such this suggests itself as a potentially fertile ground for further investigations in this field of experiential architectures embedded in the real world.

In summary, our initial experiences with the Hololens have been exciting, and many of the (long held) promises of AR/MR are fulfilled by the hardware platform. It seems clear to us that the Hololens could become an indispensable tool for a range of academic and professional uses, even in its current ``Development" incarnation. 
It is naturally anticipated that future generations of the Hololens (and future devices with similar AR/MR functionality) will only serve to enhance the utility and user experience of AR/MR. In fact, it seems to us that the base hardware is now sufficiently advanced that a fully realised nascent AR platform has potentially been created, which can move up the ``slope of enlightenment" as the hardware, software and user-base grow and mature. Based on our initial experiences, we anticipate that with this platform the many intriguing possibilities of AR/MR will soon become accessible for exploration by a wide range of researchers, across all disciplines and industries, for the full gamut of scenarios and environments.

\section*{References}
\phantomsection
\label{csl:1}Abboud, R. (2014). {Architecture in an Age of Augmented Reality: Opportunities and Obstacles for Mobile AR in Design, Construction, and Post-Completion}. \href{http://www.codessi.net/architecture-age-augmented-reality}{Retrieved from http://www.codessi.net/architecture-age-augmented-reality}

\phantomsection
\label{csl:2}Portman, M. E., Natapov, A., \& Fisher-Gewirtzman, D. (2015). {To go where no man has gone before: Virtual reality in architecture landscape architecture and environmental planning}. Computers, Environment and Urban Systems, 54, 376-384. \href{http://doi.org/10.1016/j.compenvurbsys.2015.05.001}{DOI: 10.1016/j.compenvurbsys.2015.05.001}

\phantomsection
\label{csl:3}Trimble. (2015). {Trimble HoloLens Video}. \href{https://community.trimble.com/videos/1016}{Retrieved from https://community.trimble.com/videos/1016}

\phantomsection
\label{csl:4}Microsoft. (2015). {Autodesk Maya and Microsoft HoloLens (video of conference presentation)}. \href{https://www.youtube.com/watch?v=yADhOKEbZ5Q}{Retrieved from https://www.youtube.com/watch?v=yADhOKEbZ5Q}

\phantomsection
\label{csl:5}Lynn, G., \& Erikson, S. (2016). {Architect Greg Lynn uses HoloLens, Trimble technology at Venice Biennale (video)}. \href{https://blogs.windows.com/devices/2016/05/27/architect-uses-hololens-trimble-technology-at-venice-biennale/}{Retrieved from https://blogs.windows.com/devices/2016/05/27/architect-uses-hololens-trimble-technology-at-venice-biennale/}

\phantomsection
\label{csl:6}Hockett, P., \& Ingleby, T. (2016). {Augmented Reality with Hololens: Experiential Architectures Embedded in the Real World (Figshare Repository), \href{http://dx.doi.org/10.6084/m9.figshare.c.3470907}{DOI: 10.6084/m9.figshare.c.3470907}

\phantomsection
\label{csl:7}Kroon, D.-J. (2011). {Matlab 3D figure to 3D (X)HTML (Matlab File Exchange)}. \href{https://www.mathworks.com/matlabcentral/fileexchange/32207-matlab-3d-figure-to-3d--x-html}{Retrieved from https://www.mathworks.com/matlabcentral/fileexchange/32207-matlab-3d-figure-to-3d--x-html}

\phantomsection
\label{csl:8}Microsoft. (2016). {Hololens website (Hololens.com)}. \href{http://hololens.com}{Retrieved from http://hololens.com}

\phantomsection
\label{csl:9}Hockett, P. (2016). {Hololens @femtolab.ca: week 4, basic laser lab use (video)}. \href{http://femtolab.ca/?p=712}{Retrieved from http://femtolab.ca/?p=712}

\phantomsection
\label{csl:10}Hockett, P. (2016). {Hololens @femtolab.ca: week 2, Basic data visualization and immersion (video), \href{http://dx.doi.org/10.6084/m9.figshare.3842907}{DOI: 10.6084/m9.figshare.3842907}}.

\phantomsection
\label{csl:11}Hockett, P., Bisgaard, C. Z., Clarkin, O. J., \& Stolow, A. (2011). {Time-resolved imaging of purely valence-electron dynamics during a chemical reaction}. Nat Phys, 7(8), 612-615. \href{http://doi.org/10.1038/nphys1980}{DOI: 10.1038/nphys1980}

\phantomsection
\label{csl:12}Schubert, G., Schattel, D., Tönnis, M., Klinker, G., \& Petzold, F. (2015). {Tangible Mixed Reality On-Site: Interactive Augmented Visualisations from Architectural Working Models in Urban Design}. In Communications in Computer and Information Science (pp. 55-74). Springer Science Business Media. \href{http://doi.org/10.1007/978-3-662-47386-3_4}{DOI: 10.1007/978-3-662-47386-3\_4}

\phantomsection
\label{csl:13}Hockett, P., \& Ingleby, T. (2016). {Hololens @femtolab.ca: week 3, Large Scale and In-Situ Visualizations (video), \href{http://dx.doi.org/10.6084/m9.figshare.3846672}{DOI: 10.6084/m9.figshare.3846672}}. 

\phantomsection
\label{csl:14}Gregory, R. L. (1995). {Realities of Virtual Reality}. Perception, 24(12), 1369-1371. \href{http://doi.org/10.1068/p241369}{DOI: 10.1068/p241369}

\phantomsection
\label{csl:15}Hockett, P., Lux, C., Wollenhaupt, M., \& Baumert, T. (2015). {Maximum-information photoelectron metrology}. Phys. Rev. A, 92(1). \href{http://doi.org/10.1103/physreva.92.013412}{DOI: 10.1103/physreva.92.013412}

\phantomsection
\label{csl:16}Hockett, P., Ripani, E., Rytwinski, A., \& Stolow, A. (2013). {Probing ultrafast dynamics with time-resolved multi-dimensional coincidence imaging: butadiene}. Journal of Modern Optics, 60(17), 1409-1425. \href{http://doi.org/10.1080/09500340.2013.801525}{DOI: 10.1080/09500340.2013.801525}

\phantomsection
\label{csl:17}Brekelmans, J., \& Brekel.com. (2016). {HoloLens with Autodesk MotionBuilder (video)}. \href{https://forums.hololens.com/discussion/1201/hololens-with-autodesk-motionbuilder}{Retrieved from https://forums.hololens.com/discussion/1201/hololens-with-autodesk-motionbuilder}

\phantomsection
\label{csl:18}Hockett, P., \& Ingleby, T. (2016). {Hololens @femtolab.ca: Computational Architecture at Large Scales (video), \href{http://dx.doi.org/10.6084/m9.figshare.3846675}{DOI: 10.6084/m9.figshare.3846675}}. 

\phantomsection
\label{csl:19}Kuhlen, T. W., \& Hentschel, B. (2014). {Quo Vadis CAVE: Does Immersive Visualization Still Matter?}. IEEE Comput. Grap. Appl., 34(5), 14-21. \href{http://doi.org/10.1109/mcg.2014.97}{DOI: 10.1109/mcg.2014.97}

\phantomsection
\label{csl:20}Bright, P. (2015). {The HoloLens’ limited field of view doesn’t matter, and here’s why - It's all about who you sell it to (Arstechnica)}. \href{http://arstechnica.com/gadgets/2015/07/the-hololenss-limited-field-of-view-doesnt-matter-and-heres-why/}{Retrieved from http://arstechnica.com/gadgets/2015/07/the-hololenss-limited-field-of-view-doesnt-matter-and-heres-why/}

\phantomsection
\label{csl:21}Bright, P. (2015). {HoloLens: Still magical, but with the ugly taint of reality (Arstechnica)}. \href{http://arstechnica.com/gadgets/2015/05/hololens-still-magical-but-with-the-ugly-taint-of-reality/}{Retrieved from http://arstechnica.com/gadgets/2015/05/hololens-still-magical-but-with-the-ugly-taint-of-reality/}

\phantomsection
\label{csl:22}Geng, J. (2013). {Three-dimensional display technologies}. Advances in Optics and Photonics, 5(4), 456. \href{http://doi.org/10.1364/aop.5.000456}{DOI: 10.1364/aop.5.000456}

\phantomsection
\label{csl:23}Olshannikova, E., Ometov, A., Koucheryavy, Y., \& Olsson, T. (2015). {Visualizing Big Data with augmented and virtual reality: challenges and research agenda}. Journal of Big Data, 2(1). \href{http://doi.org/10.1186/s40537-015-0031-2}{DOI: 10.1186/s40537-015-0031-2}

\phantomsection
\label{csl:24}Kwon, O.-H., Muelder, C., Lee, K., \& Ma, K.-L. (2016). {A Study of Layout Rendering, and Interaction Methods for Immersive Graph Visualization}. {IEEE} Trans. Visual. Comput. Graphics, 22(7), 1802-1815. \href{http://doi.org/10.1109/tvcg.2016.2520921}{DOI: 10.1109/tvcg.2016.2520921}

\end{document}